\documentclass[]{ceurart}

\usepackage{listings}
\usepackage{booktabs}
\usepackage{graphicx}

\begin{document}

\copyrightyear{2026}
\copyrightclause{Copyright for this paper by its authors.
  Use permitted under Creative Commons License Attribution 4.0
  International (CC BY 4.0).}

\conference{SEI-HHAI workshop at HHAI 2026, Brussels, Belgium}

\title{Beyond the Mouth: Upper-Face Affective Cues in Audiovisual Sentence
  Recognition under Acoustic Uncertainty}

\author[1]{Zhou Yang}[%
  email=zhou.yang@oulu.fi,
]
\address[1]{Faculty of Education and Psychology, University of Oulu, Finland}

\author[2]{Yueyi Yang}[%
  email=yueyi.yang@oulu.fi,
]
\address[2]{Center for Machine Vision and Signal Analysis, University of Oulu, Finland}

\begin{abstract}
Face-to-face speech comprehension is inherently multimodal: listeners process
the acoustic signal together with visible articulation, facial expression, head
motion, and other socially relevant cues. Building multimodal AI systems that
can robustly interpret speech under realistic interaction conditions therefore
requires integrating information from both speech and vision. Computational
audiovisual speech systems often emphasize the mouth region as the primary
visual source of speech information, while affective facial expression is
typically treated as a separate emotion-recognition target. This paper presents
a controlled cue-ablation study asking whether upper-face affective information
contributes to audiovisual sentence recognition beyond audio and mouth-region
cues, especially when audio is degraded. Using the CREMA-D audiovisual emotional
speech corpus, we train feature-based sentence classifiers under four cue
conditions: audio only (A), audio plus mouth/lower-face features (A+M), audio
plus upper-face features (A+U), and audio plus both mouth and upper-face
features (A+M+U). We evaluate clean audio and pink-noise conditions at +10~dB,
+5~dB, and 0~dB SNR using actor-independent splits. Results show that
mouth/lower-face features provide clear robustness benefits under degraded
audio: at 0~dB, A+M improves accuracy over A by 0.0794, with actor-bootstrap
95\% CI [0.0296, 0.1298]. Upper-face affective cues beyond the mouth show a
more qualified pattern: the direct A+M+U gain over A+M is small and uncertain,
but full-face models improve expected calibration error across SNRs and
outperform shuffled upper-face controls under noisy audio. These findings
support a cautious account in which affective full-face information contributes
to multimodal robustness and confidence estimation under acoustic uncertainty,
without implying that upper-face cues directly encode lexical content. More
broadly, the study highlights how socially expressive facial information may
support multimodal inference in human-centered audiovisual interaction systems.
\end{abstract}


\maketitle

\section{Introduction}

Speech perception in face-to-face settings is inherently multimodal. Human
listeners integrate the acoustic signal together with visible articulation,
facial expression, head motion, and other socially relevant contextual cues
while interpreting spoken language. Classic work on audiovisual speech
perception demonstrates that visible articulation can alter phonetic
perception~\cite{McGurk1976}, and speech-in-noise studies show that visual
access to the talker becomes especially important when the auditory signal is
unreliable~\cite{Sumby1954}. More recent psycholinguistic and multimodal
interaction accounts further argue that language processing in natural settings
should be understood as predictive, socially situated, and embodied rather than
as isolated acoustic decoding~\cite{Holler2019,Benetti2023}. These questions are
increasingly relevant not only for cognitive science, but also for multimodal AI
systems designed for human-centered interaction, where speech, vision, and
affective signals must often be interpreted jointly under noisy real-world
conditions.

Many computational treatments of audiovisual speech preserve only part of this
setting. Audiovisual speech recognition systems often focus primarily on the
mouth region because it contains visible articulatory information closely tied
to phonetic content. Affective-computing systems, by contrast, frequently treat
facial expression as a standalone emotion-recognition target rather than as
information that may shape language-oriented inference. While this division is
methodologically useful, it leaves open a narrower but important question for
multimodal interaction systems: when audio is uncertain, does upper-face
affective information provide useful constraints for sentence recognition beyond
mouth-region visual speech cues?

\begin{figure}[t]
    \centering
    \includegraphics[width=\linewidth]{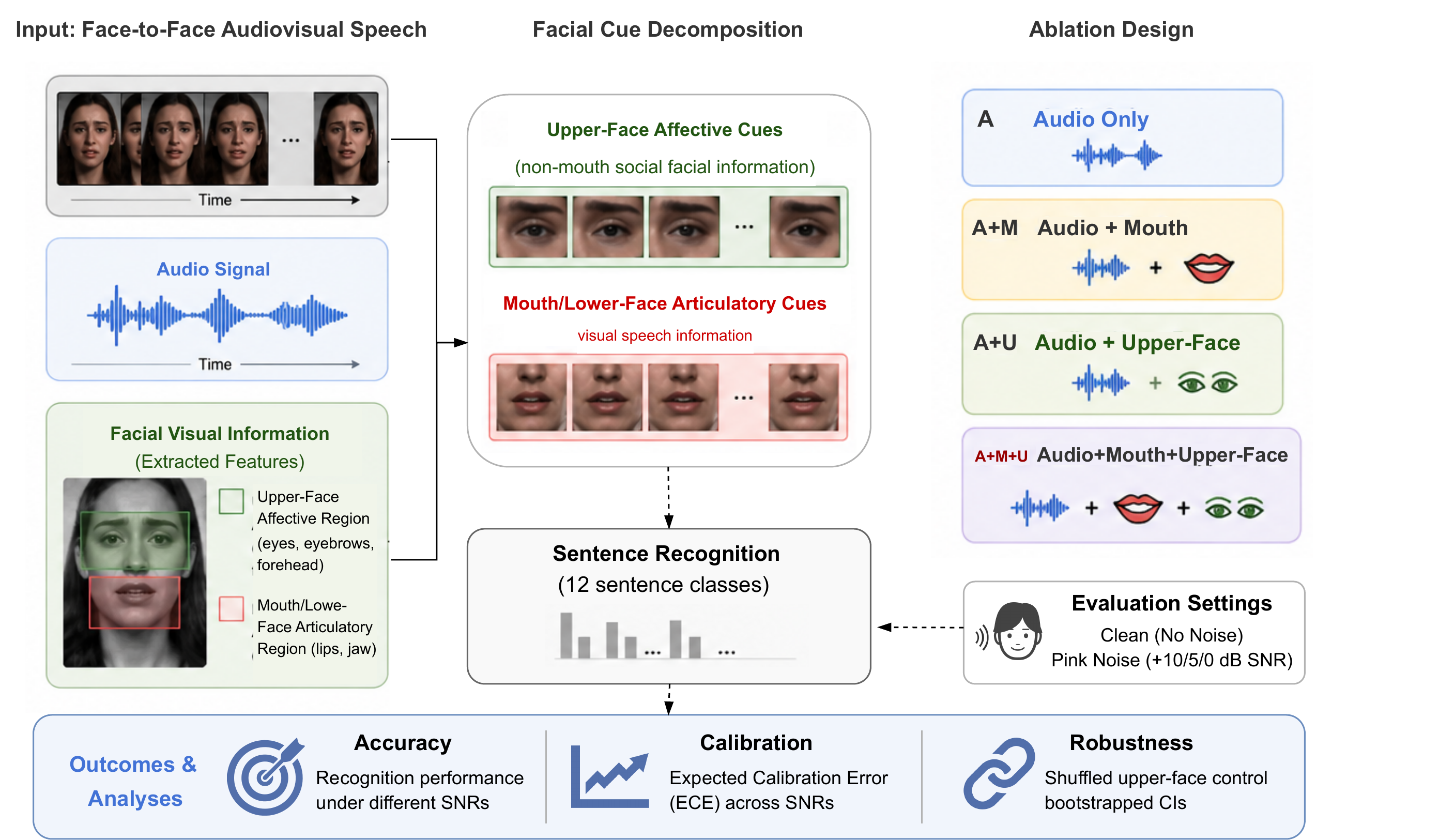}
    \caption{
Overview of the proposed multimodal cue-ablation framework for audiovisual
sentence recognition under acoustic uncertainty. The framework decomposes
facial information into mouth/lower-face articulatory cues and upper-face
affective cues, and evaluates their contributions under four cue conditions
(A, A+M, A+U, A+M+U) across multiple noise levels. The study analyzes not
only sentence-recognition accuracy, but also calibration and
alignment-sensitive robustness under degraded audio conditions.
    }
    \label{fig:framework}
\end{figure}

This paper addresses that question through a controlled computational
cue-ablation study. Rather than building a large end-to-end audiovisual speech
recognizer, we use interpretable clip-level features extracted from a public
audiovisual emotional speech dataset. This design allows us to isolate how
different facial regions contribute to multimodal inference under acoustic
uncertainty, which is often difficult to observe in large black-box models. An overview of the proposed multimodal cue-ablation framework is shown in
Figure~\ref{fig:framework}.

The
goal is not to claim that eyebrows, eye movements, or upper-face expressions
directly encode lexical content. Instead, the goal is more modest and testable:
to examine whether aligned full-face affective information contributes to
robustness, calibration, or cue integration when a model must identify spoken
sentences under degraded audio conditions.

The task uses CREMA-D, a corpus of actors producing short sentences with varied
emotional expressions~\cite{Cao2014}. CREMA-D is well suited to this pilot
study because it systematically combines sentence identity with emotionally
expressive facial behavior and includes enough speakers for actor-independent
evaluation. We treat closed-set sentence identification among 12 sentence
categories as a simplified proxy for audiovisual sentence recognition rather
than as open-ended automatic speech recognition or full semantic comprehension.

The main contributions of this paper are as follows:
\begin{itemize}
    \item We operationalize a question about socially situated audiovisual
    language perception as a controlled computational cue-ablation problem,
    bridging affective multimodal processing with audiovisual speech inference.

    \item We disentangle mouth/lower-face articulatory information from
    upper-face affective information through region-specific facial feature
    conditions, rather than treating the face as a single undifferentiated
    visual modality.

    \item We evaluate not only recognition accuracy but also calibration and
    robustness under acoustic uncertainty, including shuffled upper-face
    controls and actor-bootstrap uncertainty estimates for cautious
    interpretation of multimodal cue effects.
\end{itemize}

\section{Related Work}

\subsection{Multimodal audiovisual speech perception under acoustic uncertainty}

Audiovisual speech perception research has long shown that visible articulation
can influence what listeners perceive. The McGurk effect is a canonical
demonstration that auditory and visual speech cues are integrated rather than
processed independently~\cite{McGurk1976}. A complementary line of work on
speech intelligibility in noise shows that access to the speaker's face and lips
can improve speech recognition when acoustic information is
degraded~\cite{Sumby1954}. These findings motivate multimodal computational
systems that integrate speech and vision, particularly under noisy or ambiguous
interaction conditions.

However, much audiovisual speech work privileges the mouth region. This is
sensible because lip shape, lip aperture, jaw movement, and visible articulatory
gestures are closely related to phonetic information. Yet in natural face-to-face
interaction listeners observe the whole face rather than only the mouth.
Non-mouth facial cues may convey affective state, attention, emphasis, stance,
or interactional context that can shape interpretation even when they are not
phonetic cues in the strict sense.

\subsection{Affective facial information in multimodal human communication}

Recent psycholinguistic work argues that face-to-face language processing
involves temporally offset visual and vocal signals that must be integrated
under tight interactional constraints~\cite{Holler2019}. Benetti, Ferrari, and
Pavani further propose that multimodal face-to-face communication involves
multiplex signals, multimodal gestalts, and multilevel predictions, linking
psycholinguistic theory to sensory-neuroscience mechanisms~\cite{Benetti2023}.
In this view, visible facial information is not merely decorative context but
part of the perceptual environment in which spoken meaning is inferred.

Within affective computing, facial expression is commonly modeled as a target
for emotion recognition. While effective for emotion classification, this
framing can separate affective perception from language-oriented inference. For
multimodal human-AI interaction systems, a broader question is whether
affective facial information can support robustness, uncertainty estimation, or
cue integration when speech signals are degraded or incomplete. The present
study focuses specifically on affective cues because emotional audiovisual
speech corpora provide a controlled setting for studying socially expressive
facial behavior, while acknowledging that natural social interaction also
involves other signals such as intention, attention, and interpersonal stance.

\subsection{Full-face multimodal modeling and emotional audiovisual speech}

The present study is closely related to emotional lip-reading and recent
full-face audiovisual language modeling. EMOLIPS, for example, used CREMA-D and
RAVDESS to investigate emotional speech lip-reading, showing that emotion-aware
visual modeling can improve phrase recognition~\cite{Ryumin2023}. Recent
emotion-aware audiovisual language modeling has also incorporated full-face
visual cues into speech-language modeling for expressive generation and
emotion-related tasks~\cite{Tan2025}. These studies suggest that emotionally
expressive facial information can be computationally useful beyond traditional
mouth-centered speech modeling.

Our contribution differs in emphasis. We do not primarily optimize emotional
lip-reading or expressive generation. Instead, we use a controlled feature-based
ablation framework to isolate whether upper-face affective information provides
additional value beyond mouth-region visual articulation in a language-oriented
sentence-recognition task. We further examine whether such effects emerge not
only in recognition accuracy, but also in calibration, log-loss, and
alignment-sensitive robustness under degraded audio conditions. This narrower
and more cautious framing is important because upper-face affective cues should
not be interpreted as directly encoding lexical content.

\section{Data and Task}

\subsection{Dataset}

We use CREMA-D, the Crowd-sourced Emotional Multimodal Actors
Dataset~\cite{Cao2014}. CREMA-D contains short audiovisual recordings of actors
producing sentence materials with different emotional expressions. In the
extracted manifest used for this study, the dataset contained 7,442 audiovisual
clips from 91 actors. The clips covered 12 sentence categories and six emotion
categories: anger, disgust, fear, happy, neutral, and sad.

All 7,442 clips had corresponding WAV audio and video files. One clip,
\texttt{1076\_MTI\_SAD\_XX}, was excluded from the final audiovisual feature
table because the face tracker did not detect a face. The final table used for
audiovisual modeling therefore contained 7,441 clips.

\subsection{Prediction target and evaluation split}

The prediction target is sentence identity among the 12 CREMA-D sentence
categories. This is a closed-set sentence classification task. It should not be
interpreted as open-ended ASR, dialogue understanding, or full language
comprehension; it is a controlled proxy that lets us test whether different
audiovisual cue groups help recover which sentence was spoken.

To reduce speaker-identity leakage, actors rather than individual clips were
assigned to train, validation, and test sets. Hyperparameters were selected on
the validation actors, and final metrics were computed on held-out test actors.
Table~\ref{tab:split} summarizes the split.

\begin{table}[h]
  \caption{Actor-independent dataset split used for model selection and final evaluation.}
  \label{tab:split}
  \begin{tabular}{lrrcc}
    \toprule
    Split      & Actors & Clips & Sentences & Emotions \\
    \midrule
    Train      & 63     & 5,147 & 12        & 6        \\
    Validation & 14     & 1,148 & 12        & 6        \\
    Test       & 14     & 1,147 & 12        & 6        \\
    \bottomrule
  \end{tabular}
\end{table}

\section{Methods}

\subsection{Audio degradation}

We evaluated four audio conditions: clean audio, pink noise at +10~dB SNR, pink
noise at +5~dB SNR, and pink noise at 0~dB SNR. Noise was generated
deterministically in memory from each clean WAV file using a fixed seed policy.
Visual features were extracted once from the original videos and reused across
audio conditions. This design tests whether visual cue benefits increase as the
acoustic signal becomes less reliable.

\subsection{Feature extraction}

Audio features were extracted with openSMILE using the eGeMAPSv02 functionals
feature set~\cite{Eyben2010,Eyben2016}. These clip-level features capture
prosodic, spectral, and voice-quality information commonly used in speech and
affective-speech analysis.

Facial features were extracted with MediaPipe Face Landmarker~\cite{MediaPipe},
which outputs 3D facial landmarks, facial blendshape scores, and transformation
information. We separated visual features into two groups. \emph{Mouth/lower-face
features} included lip aperture, mouth width, jaw-to-nose distance, mouth opening
ratio, mouth-related blendshape scores, and temporal summary statistics.
\emph{Upper-face/non-mouth features} included non-mouth blendshape scores and
head-pose-related transformation features. For efficiency, visual features were
extracted from every second video frame. Frame-level features were aggregated to
clip-level summaries using mean, standard deviation, minimum, maximum,
percentiles, and mean absolute derivative.

\subsection{Cue conditions and shuffled control}

Separate models were trained for four cue conditions: audio only (A), audio plus
mouth/lower-face features (A+M), audio plus upper-face/non-mouth features (A+U),
and audio plus both mouth and upper-face features (A+M+U). The main theoretical
contrast is A+M+U versus A+M, because this asks whether non-mouth facial
information adds value after visible articulation is already available.

We also implemented a \emph{shuffled upper-face control}. In this control,
upper-face features were mismatched across clips within each split, while audio
and mouth features remained aligned with the original clip. This preserves the
number and type of upper-face features while disrupting clip-level alignment. The
control tests whether any full-face benefit reflects aligned facial information
rather than feature-count inflation, actor-specific patterns, or recording
artifacts.

\subsection{Model, metrics, and uncertainty estimates}

The baseline model is multinomial logistic regression with standardized features
and median imputation, implemented with scikit-learn~\cite{Pedregosa2011}. The
regularization strength $C$ was selected on the validation set from a small grid:
0.001, 0.01, 0.1, 1.0, and 10.0. Separate models were trained for each audio
condition and cue condition. This model family was chosen because the goal is an
interpretable workshop pilot rather than a benchmark-optimized audiovisual speech
recognizer.

Performance was evaluated using accuracy, macro-F1, log loss, and expected
calibration error (ECE). Accuracy and macro-F1 evaluate sentence classification.
Log loss measures the probability assigned to the correct class. ECE measures
whether model confidence is well calibrated; lower ECE indicates better
calibration~\cite{Guo2017}. For key contrasts, we computed actor-level bootstrap
confidence intervals over the 14 held-out test actors using 1,000 bootstrap
draws. Actor-level resampling preserves the speaker-level dependence structure
more appropriately than treating all clips as independent.

\section{Results}

\subsection{Main aligned cue-ablation results}

Table~\ref{tab:main} shows the aligned cue-ablation results across audio
conditions. Audio-only accuracy decreased as noise increased, from 0.8325 in
clean audio to 0.6344 at 0~dB. This confirms that the degradation manipulation
made the auditory signal less reliable.

\begin{table*}
  \caption{Main aligned cue-ablation results by SNR condition. A = audio;
    M = mouth/lower-face features; U = upper-face/non-mouth features.
    Lower log loss and lower ECE are better.}
  \label{tab:main}
  \begin{tabular}{llcccc}
    \toprule
    SNR    & Condition & Accuracy & Macro-F1 & Log loss & ECE    \\
    \midrule
    Clean  & A         & 0.8325   & 0.8313   & 0.5559   & 0.0316 \\
    Clean  & A+M       & 0.8377   & 0.8282   & 0.5846   & 0.0250 \\
    Clean  & A+U       & 0.8124   & 0.8065   & 0.6098   & 0.0550 \\
    Clean  & A+M+U     & 0.8264   & 0.8168   & 0.6211   & 0.0130 \\
    \midrule
    +10~dB & A         & 0.7993   & 0.7931   & 0.6270   & 0.0410 \\
    +10~dB & A+M       & 0.8115   & 0.7966   & 0.6413   & 0.0345 \\
    +10~dB & A+U       & 0.7836   & 0.7745   & 0.6738   & 0.0671 \\
    +10~dB & A+M+U     & 0.8063   & 0.7931   & 0.6613   & 0.0187 \\
    \midrule
    +5~dB  & A         & 0.7452   & 0.7344   & 0.8060   & 0.0312 \\
    +5~dB  & A+M       & 0.7714   & 0.7530   & 0.7470   & 0.0327 \\
    +5~dB  & A+U       & 0.7321   & 0.7173   & 0.8025   & 0.0481 \\
    +5~dB  & A+M+U     & 0.7696   & 0.7537   & 0.7650   & 0.0268 \\
    \midrule
    0~dB   & A         & 0.6344   & 0.6162   & 1.1198   & 0.0306 \\
    0~dB   & A+M       & 0.7138   & 0.6913   & 0.9204   & 0.0302 \\
    0~dB   & A+U       & 0.6370   & 0.6142   & 1.0638   & 0.0341 \\
    0~dB   & A+M+U     & 0.7243   & 0.7041   & 0.9246   & 0.0266 \\
    \bottomrule
  \end{tabular}
\end{table*}

\subsection{Mouth/lower-face visual features under noise}

Mouth/lower-face features provided the clearest robustness benefit. The accuracy
gain of A+M over A increased as audio became noisier: $+0.0052$ in clean audio,
$+0.0122$ at $+10$~dB, $+0.0262$ at $+5$~dB, and $+0.0794$ at 0~dB. At 0~dB,
the actor-level bootstrap supported this effect: accuracy difference $=+0.0794$,
95\% CI $[{+0.0296},{+0.1298}]$. Log loss also improved at 0~dB: difference
$=-0.2007$, 95\% CI $[-0.3764,-0.0207]$. Thus, the experiment supports the basic
audiovisual speech-recognition prediction that visible mouth cues help most when
audio is uncertain.

\subsection{Upper-face/full-face effects beyond the mouth}

The evidence for upper-face/full-face cues beyond mouth features was more
qualified. In clean audio, A+M+U did not improve accuracy over A+M. At $+10$~dB
and $+5$~dB, A+M+U was also slightly below A+M in accuracy. At the hardest noise
level, 0~dB, A+M+U reached accuracy 0.7243 compared with 0.7138 for A+M, a gain
of $+0.0105$. However, the actor-bootstrap confidence interval for this direct
contrast crossed zero: 95\% CI $[-0.0052, {+0.0254}]$.

Calibration told a more favorable story for full-face information. Adding
upper-face/full-face features to A+M reduced ECE in all audio conditions:
$-0.0120$ in clean audio, $-0.0158$ at $+10$~dB, $-0.0060$ at $+5$~dB, and
$-0.0037$ at 0~dB. Because lower ECE indicates better calibration, this suggests
that full-face information may help probability estimates even when it does not
substantially improve top-1 accuracy.

\subsection{Shuffled upper-face control}

The shuffled upper-face control provides the strongest evidence that aligned
upper-face information is not merely extra feature quantity. Under degraded audio,
aligned A+M+U consistently outperformed shuffled A+M+U. At $+10$~dB, the aligned
model exceeded the shuffled control by $+0.0183$ accuracy points, 95\% CI
$[{+0.0026},{+0.0366}]$. At $+5$~dB, the difference was $+0.0253$, 95\% CI
$[{+0.0044},{+0.0488}]$. At 0~dB, the difference increased to $+0.0305$, 95\%
CI $[{+0.0113},{+0.0480}]$. At 0~dB, aligned A+M+U also improved log loss
relative to shuffled A+M+U: difference $= -0.0889$, 95\% CI
$[-0.1166, -0.0602]$.

\begin{figure}[t]
  \centering
  \includegraphics[width=\linewidth]{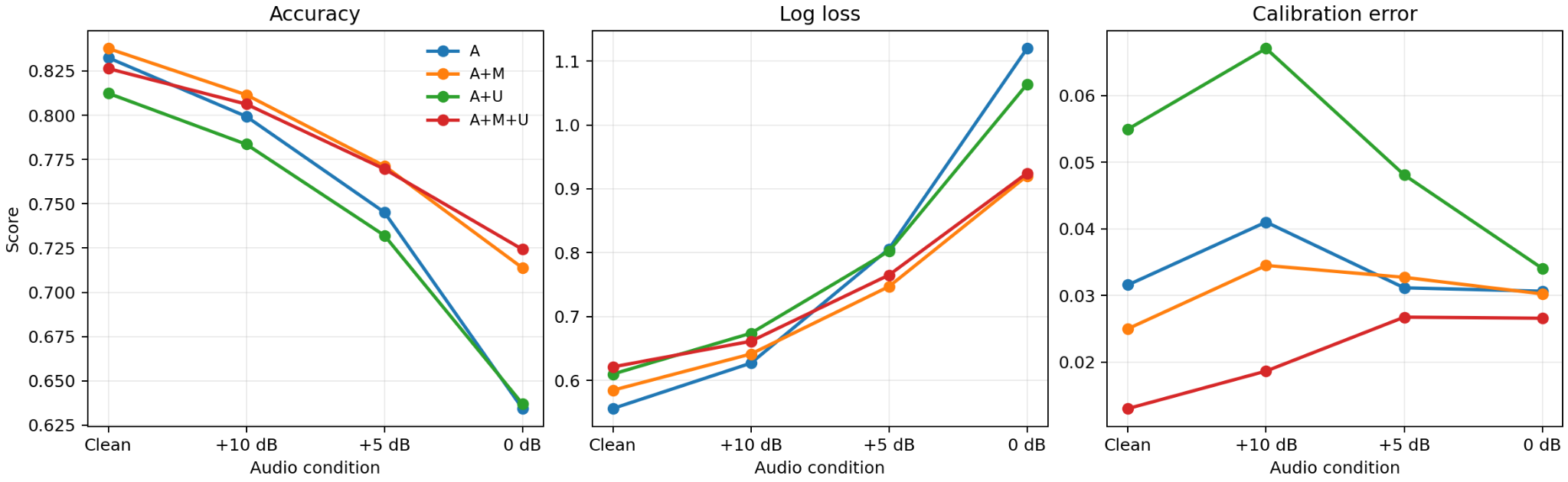}
  \caption{Accuracy, log loss, and expected calibration error across audio
    conditions and cue-ablation settings. Visual features become more important
    as audio is degraded; full-face models show their clearest advantage in
    calibration and shuffled-control comparisons rather than in direct top-1
    gains over mouth-only input.}
  \label{fig:results}
\end{figure}

\begin{table}[h]
  \caption{Key actor-bootstrap contrasts. Confidence intervals are based on
    1,000 resamples over held-out test actors.}
  \label{tab:bootstrap}
  \begin{tabular}{llccc}
    \toprule
    SNR    & Contrast                             & Metric   & Estimate  & 95\% CI              \\
    \midrule
    0~dB   & A+M minus A                          & Accuracy & $+0.0794$ & $[+0.0296, +0.1298]$ \\
    0~dB   & A+M minus A                          & Log loss & $-0.2007$ & $[-0.3764, -0.0207]$ \\
    0~dB   & A+M+U minus A+M                      & Accuracy & $+0.0105$ & $[-0.0052, +0.0254]$ \\
    0~dB   & A+M+U minus A+M                      & ECE      & $-0.0037$ & $[-0.0195, +0.0216]$ \\
    +10~dB & Aligned A+M+U minus shuffled A+M+U   & Accuracy & $+0.0183$ & $[+0.0026, +0.0366]$ \\
    +5~dB  & Aligned A+M+U minus shuffled A+M+U   & Accuracy & $+0.0253$ & $[+0.0044, +0.0488]$ \\
    0~dB   & Aligned A+M+U minus shuffled A+M+U   & Accuracy & $+0.0305$ & $[+0.0113, +0.0480]$ \\
    0~dB   & Aligned A+M+U minus shuffled A+M+U   & Log loss & $-0.0889$ & $[-0.1166, -0.0602]$ \\
    \bottomrule
  \end{tabular}
\end{table}

\section{Discussion}

\subsection{Multimodal cue integration under acoustic uncertainty}

This pilot study supports a differentiated account of multimodal audiovisual cue
integration. Mouth/lower-face features behave consistently with established
audiovisual speech perception findings: they improve sentence recognition, and
their benefit becomes larger as the acoustic signal becomes less reliable. This
result validates the experimental setup and confirms that the closed-set
CREMA-D task is sensitive to visually grounded speech information.

The upper-face/full-face result is more nuanced. Upper-face affective features
did not produce a strong direct top-1 accuracy gain over mouth-only features
across all conditions. This is theoretically expected because upper-face
movements are not articulatory cues in the same way as lip and jaw movements.
They should therefore not be interpreted as directly encoding lexical content.
However, full-face features improved calibration across SNR levels and showed
alignment-sensitive effects in the shuffled-control condition. When upper-face
information was mismatched across clips, full-face performance dropped under
degraded audio, suggesting that aligned affective facial information contributes
to multimodal inference in a nontrivial way.

\subsection{Affective facial information as interaction-relevant context}

The most cautious interpretation is therefore not that upper-face affective
signals carry sentence identity themselves. Rather, upper-face and full-face
affective information may provide uncertainty-relevant contextual cues that help
stabilize multimodal inference when acoustic information is incomplete. This
interpretation is consistent with socially situated accounts of language
processing, where spoken understanding emerges through the coordination of
acoustic speech, visible articulation, affective expression, and broader
interactional context~\cite{Holler2019,Benetti2023}.

From a multimodal human-AI interaction perspective, these findings suggest that
socially expressive facial information may contribute not only to explicit
emotion recognition, but also to robust and calibrated language-oriented
processing under noisy real-world conditions. Importantly, affective expression
represents only one component of social interaction; natural communication also
involves intention, attention, interpersonal stance, and other contextual
signals that are not captured in the present dataset.

The shuffled-control results are particularly informative for interpreting the
role of upper-face affective cues. If the study had only compared A+M+U to A,
the improvement under noise could largely be attributed to mouth-region visual
speech information. Conversely, if it had only compared A+M+U to A+M, the
full-face effect would appear weak. The shuffled-control analysis reveals a more
subtle pattern: aligned upper-face affective information matters under acoustic
uncertainty, but its contribution emerges more clearly in robustness and
calibration behavior than as a large direct top-1 accuracy gain within the
present linear baseline.

\subsection{Implications for multimodal human-AI interaction}

For multimodal human-AI interaction systems, the results motivate a cautious
design principle. Systems that process face-to-face speech should not treat the
mouth as the only relevant visual channel, but they should also avoid treating
facial affect as a direct lexical signal. Instead, upper-face and full-face
affective information may be most useful for uncertainty estimation, robustness
under degraded audio, calibration, or broader interaction-relevant inference
under noisy conditions. This distinction helps connect audiovisual speech
recognition with affective multimodal processing without collapsing them into
the same computational task.

\section{Limitations and Future Work}

First, the task is closed-set sentence identification among 12 fixed sentences.
It is not open-ended ASR, dialogue understanding, or semantic comprehension. The
results should therefore be interpreted as a controlled proxy for audiovisual
sentence recognition under acoustic uncertainty rather than as a complete model
of multimodal language understanding.

Second, the current study operates at the feature level rather than through
end-to-end audiovisual representation learning. We use clip-level facial and
audio features together with linear classifiers instead of raw video-based
computer vision pipelines. This design improves interpretability, reproducibility,
and controlled cue analysis, but it may miss temporal or nonlinear interactions
between audio, mouth motion, and affective facial expression. Future work could
extend the present framework with lightweight temporal fusion models or
end-to-end multimodal architectures while preserving interpretable cue-ablation
analysis.

Third, upper-face features may contain actor-specific, pose-related, or
recording-related information. The actor-independent split and shuffled-control
analysis reduce this risk, but they do not eliminate all possible confounds.
Additional controls could include feature-family permutation importance,
emotion-balanced error analyses, and replication on datasets such as RAVDESS or
MEAD.

Fourth, visual features were extracted using MediaPipe Face Landmarker rather
than a specialized facial action unit framework such as OpenFace. MediaPipe was
chosen for operational simplicity, robustness, and reproducibility. Only one
clip failed face tracking, but future work could compare MediaPipe blendshape
features with action-unit representations or learned facial embeddings.

Finally, the present analysis uses CREMA-D only. CREMA-D is valuable because it
systematically combines sentence identity with emotional facial expression, but
it remains an acted short-utterance corpus rather than natural conversational
interaction. Future work should examine whether the calibration and
alignment-sensitive effects observed here generalize to more naturalistic
multimodal human-AI interaction settings involving spontaneous audiovisual
communication, richer social context, and broader interactional signals beyond
affective expression alone.

\section{Conclusion}

This paper presented a controlled computational cue-ablation study of
upper-face/full-face affective information in audiovisual sentence recognition
under acoustic uncertainty. Using the CREMA-D emotional audiovisual speech
corpus, we showed that mouth/lower-face visual features provide clear robustness
benefits when audio quality degrades. Upper-face/full-face affective cues
exhibited a more nuanced but theoretically meaningful pattern: direct gains over
mouth-only features were modest, yet full-face models improved calibration and
outperformed shuffled upper-face controls under noisy conditions. These findings
support a cautious interpretation in which affective facial information
contributes to multimodal robustness and uncertainty-sensitive inference during
face-to-face language processing, without implying that upper-face cues directly
encode lexical content. More broadly, the study highlights the potential role of
socially expressive facial information in multimodal human-AI interaction
systems operating under realistic acoustic uncertainty.


\newpage

\section*{Declaration on Generative AI}
Generative AI (ChatGPT) were used exclusively to improve the clarity and fluency of English writing. They were not involved in research ideation, experimental design, data analysis, or interpretation. The authors take full responsibility for all content.

\bibliography{paper}

@article{McGurk1976,
  author  = {McGurk, Harry and MacDonald, John},
  title   = {Hearing lips and seeing voices},
  journal = {Nature},
  volume  = {264},
  pages   = {746--748},
  year    = {1976},
  doi     = {10.1038/264746a0}
}

@article{Sumby1954,
  author  = {Sumby, W. H. and Pollack, Irwin},
  title   = {Visual contribution to speech intelligibility in noise},
  journal = {Journal of the Acoustical Society of America},
  volume  = {26},
  number  = {2},
  pages   = {212--215},
  year    = {1954},
  doi     = {10.1121/1.1907309}
}

@article{Holler2019,
  author  = {Holler, Judith and Levinson, Stephen C.},
  title   = {Multimodal language processing in human communication},
  journal = {Trends in Cognitive Sciences},
  volume  = {23},
  number  = {8},
  pages   = {639--652},
  year    = {2019},
  doi     = {10.1016/j.tics.2019.05.006}
}

@article{Benetti2023,
  author  = {Benetti, Stefano and Ferrari, Anna and Pavani, Francesco},
  title   = {Multimodal processing in face-to-face interactions: {A} bridging
             link between psycholinguistics and sensory neuroscience},
  journal = {Frontiers in Human Neuroscience},
  volume  = {17},
  pages   = {1108354},
  year    = {2023},
  doi     = {10.3389/fnhum.2023.1108354}
}

@article{Cao2014,
  author  = {Cao, Houwei and Cooper, David G. and Keutmann, Michael K. and
             Gur, Ruben C. and Nenkova, Ani and Verma, Ragini},
  title   = {{CREMA-D}: Crowd-sourced Emotional Multimodal Actors Dataset},
  journal = {IEEE Transactions on Affective Computing},
  volume  = {5},
  number  = {4},
  pages   = {377--390},
  year    = {2014},
  doi     = {10.1109/TAFFC.2014.2336244}
}

@article{Ryumin2023,
  author  = {Ryumin, Dmitry and Ryumina, Elena and Ivanko, Denis},
  title   = {{EMOLIPS}: Towards reliable emotional speech lip-reading},
  journal = {Mathematics},
  volume  = {11},
  number  = {23},
  pages   = {4787},
  year    = {2023},
  doi     = {10.3390/math11234787}
}

@inproceedings{Tan2025,
  author    = {Tan, Wenda and Lian, Jiachen and Inaguma, Hirofumi and
               Tomasello, Peppe and Koehn, Philipp and Ma, Xutai},
  title     = {Seeing is Believing: Emotion-Aware Audio-Visual Language Modeling
               for Expressive Speech Generation},
  booktitle = {Findings of the Association for Computational Linguistics: EMNLP 2025},
  year      = {2025},
  url       = {https://aclanthology.org/2025.findings-emnlp.140/}
}

@inproceedings{Eyben2010,
  author    = {Eyben, Florian and W{\"o}llmer, Martin and Schuller, Bj{\"o}rn},
  title     = {{openSMILE}: The {M}unich versatile and fast open-source audio
               feature extractor},
  booktitle = {Proceedings of the 18th ACM International Conference on Multimedia},
  pages     = {1459--1462},
  year      = {2010},
  doi       = {10.1145/1873951.1874246}
}

@article{Eyben2016,
  author  = {Eyben, Florian and Scherer, Klaus R. and Schuller, Bj{\"o}rn W. and
             Sundberg, Johan and Andr{\'e}, Elisabeth and Busso, Carlos and
             Devillers, Laurence Y. and Epps, Julien and Laukka, Petri and
             Narayanan, Shrikanth S. and Truong, Khiet P.},
  title   = {The {G}eneva Minimalistic Acoustic Parameter Set ({GeMAPS}) for
             voice research and affective computing},
  journal = {IEEE Transactions on Affective Computing},
  volume  = {7},
  number  = {2},
  pages   = {190--202},
  year    = {2016},
  doi     = {10.1109/TAFFC.2015.2457417}
}

@misc{MediaPipe,
  author       = {{Google AI Edge}},
  title        = {{MediaPipe} Face Landmarker},
  howpublished = {\url{https://ai.google.dev/edge/mediapipe/solutions/vision/face_landmarker}},
  year         = {2024}
}

@inproceedings{Guo2017,
  author    = {Guo, Chuan and Pleiss, Geoff and Sun, Yu and Weinberger, Kilian Q.},
  title     = {On calibration of modern neural networks},
  booktitle = {Proceedings of the 34th International Conference on Machine Learning},
  series    = {PMLR},
  volume    = {70},
  pages     = {1321--1330},
  year      = {2017},
  url       = {https://proceedings.mlr.press/v70/guo17a.html}
}

@article{Pedregosa2011,
  author  = {Pedregosa, Fabian and Varoquaux, Ga{\"e}l and Gramfort, Alexandre and
             Michel, Vincent and Thirion, Bertrand and Grisel, Olivier and
             Blondel, Mathieu and Prettenhofer, Peter and Weiss, Ron and
             Dubourg, Vincent and Vanderplas, Jake and Passos, Alexandre and
             Cournapeau, David and Brucher, Matthieu and Perrot, Matthieu and
             Duchesnay, {\'E}douard},
  title   = {Scikit-learn: Machine Learning in {P}ython},
  journal = {Journal of Machine Learning Research},
  volume  = {12},
  pages   = {2825--2830},
  year    = {2011},
  url     = {https://jmlr.org/papers/v12/pedregosa11a.html}
}

\end{document}